\documentclass[aps,prl,twocolumn,superscriptaddress]{revtex4}
\usepackage{graphicx}
\usepackage{mathrsfs}
\usepackage{bm}
\usepackage{multirow}
\usepackage{amsmath}
\usepackage{dcolumn}
\usepackage{epstopdf}
\usepackage{dsfont}
\usepackage{amssymb}
\usepackage{tabularx}
\usepackage{array}
\usepackage{float}
\usepackage{siunitx}
\usepackage{colordvi}
\usepackage{color}
\usepackage{booktabs}
\usepackage{threeparttable}
\begin{document}
\title{Quantum Anomalous Hall Effect by Coupling Heavy Atomic Layers with CrI$_{3}$}
\author{Majeed Ur Rehman}
\affiliation{ICQD, Hefei National Laboratory for Physical Sciences at Microscale, CAS Key Laboratory of Strongly-Coupled Quantum Matter Physics, and Department of Physics, University of Science and Technology of China, Hefei, Anhui 230026, China.}
\author{Xinlong Dong}
\affiliation{College of Physics and Information Engineering, and Research Institute of Materials Science, Shanxi Normal University, Linfen, Shanxi 041004, China.}
\affiliation{ICQD, Hefei National Laboratory for Physical Sciences at Microscale, CAS Key Laboratory of Strongly-Coupled Quantum Matter Physics, and Department of Physics, University of Science and Technology of China, Hefei, Anhui 230026, China.}
\author{Tao Hou}
\affiliation{ICQD, Hefei National Laboratory for Physical Sciences at Microscale, CAS Key Laboratory of Strongly-Coupled Quantum Matter Physics, and Department of Physics, University of Science and Technology of China, Hefei, Anhui 230026, China.}
\author{Zeyu Li}
\affiliation{ICQD, Hefei National Laboratory for Physical Sciences at Microscale, CAS Key Laboratory of Strongly-Coupled Quantum Matter Physics, and Department of Physics, University of Science and Technology of China, Hefei, Anhui 230026, China.}
\author{Shifei Qi}
\affiliation{Department of Physics, Hebei Normal University, Shijiazhuang, Hebei 050024, China.}
\affiliation{ICQD, Hefei National Laboratory for Physical Sciences at Microscale, CAS Key Laboratory of Strongly-Coupled Quantum Matter Physics, and Department of Physics, University of Science and Technology of China, Hefei, Anhui 230026, China.}
\author{Zhenhua Qiao}
\email[Correspondence author:~~]{qiao@ustc.edu.cn}
\affiliation{ICQD, Hefei National Laboratory for Physical Sciences at Microscale, CAS Key Laboratory of Strongly-Coupled Quantum Matter Physics, and Department of Physics, University of Science and Technology of China, Hefei, Anhui 230026, China.}
\date{\today}

\begin{abstract}
  We explored the possibility of realizing quantum anomalous Hall effect by placing heavy-element atomic layer on top of monolayer CrI$_{3}$ with a natural cleavage surface and broken time-reversal symmetry. We showed that CrI$_{3}$/X (X = Bi, Sb, or As) systems can open up a sizable bulk gap to harbour quantum anomalous Hall effect, e.g., CrI$_{3}$/Bi is a natural magnetic insulator with a bulk gap of 30~meV, which can be further enlarged via strain engineering or adjusting spin orientations. We also found that the ferromagnetic properties (magnetic anisotropic energy and Curie temperature) of pristine CrI$_{3}$ can be further improved due to the presence of heavy atomic layers, and the spin orientation can be utilized as a useful knob to tune the band structure and Fermi level of CrI$_{3}$/Bi system. The topological nature, together with the enhanced ferromagnetism, can unlock new potential applications for CrI$_{3}$-based materials in spintronics and electronics.
\end{abstract}

\maketitle
\textit{Introduction---.} The interplay of symmetry, spin-orbit coupling, and magnetic structure together helps realize various topological phases ranging from quantum Hall effect to topological superconductor. The quantum anomalous Hall effect or magnetic topological insulator corresponds to the quantum Hall effect without applying external magnetic field~\cite{weng2015quantum,ren2016topological,he2018topological}, and
exhibits immense application potential in dissipationless quantum electronics~\cite{liu2018intrinsic,ni2018intrinsic,zanolli2018hybrid,ren2018quantum,deng2018quantum}. Typically, an element possessing high spin-orbit coupling is usually magnetically inactive, and vice versa. Therefore, the magnetic
topological insulator rarely exists in natural materials. Theoretically, many attempts have been implemented to make
it realistic in a single system by doping/adsorption~\cite{tse2011quantum,qiao2010quantum,ding2011engineering,zhang2012electrically,qiao2012microscopic,jiang2012quantum,kou2013interplay,pan2014valley,qi2016high,deng2017realization,deng2018quantum},
chemical functionalization~\cite{jin2015quantum,hsu2017quantum}, and heterostructure schemes~\cite{luo2013massive,yang2013proximity,qiao2014quantum,xu2015quantum,zhang2018strong}. However, thus far, experimentally, the quantum anomalous Hall effect has been only observed in magnetic element-doped topological insulators at very low temperatures~\cite{chang2013experimental,liu2016quantum,bestwick2015precise}. It is noteworthy that a new temperature record of 3.5 K for observing the quantum anomalous Hall effect was reported in MnBi$_2$Te$_4$ system~\cite{MnBiTe}.

It has been proposed~\cite{garrity2013chern} that magnetic topological insulators may be realized by depositing heavy-element layers on top of the surface of non-van der Waals (non-vdW) magnetic insulators. However, such a strategy has several experimental complications, such as surface
reconstruction due to existence of dangling bonds on the surface of substrates. Owing to the presence of strong chemical bonding between constituent layers, it is experimentally not easy to downsize the vertical dimension of non-vdW three-dimensional (3D) ferromagnetic/anti-ferromagnetic insulators while preserving the lateral surface geometry (or only changing it
slightly). To push away all these practical barriers and make the ideal environment more realistic, here, we suggest a versatile
experimental welcoming platform based on 2D vdW ferromagnetic insulators as a substrate for the deposition of heavy spin-orbit coupling components. Unlike a non-vdW substrate, 2D vdW ferromagnetic insulators, in principle, possess a natural cleavage plane with a perfect surface geometry free from dangling bonds, have no surface reform complication, are free of spin alignments, and are more accessible to being synthesized experimentally with excellent stability. This may attract the experimental community to invest their efforts in probing magnetic topological insulators. In addition, we address the following key queries to advance new, and possibly superior, routes for magnetic topological
insulators. (i) Can the addition of spin-orbit coupling drive trivial 2D vdW ferromagnetic systems to become large-bandgap magnetic topological insulators in a fundamental and controlled way? (ii) Can they improve the ferromagnetic properties of 2D vdW ferromagnetic insulators, such as the Curie temperature in these systems?

\begin{figure}
  \includegraphics[width=6cm,angle=0]{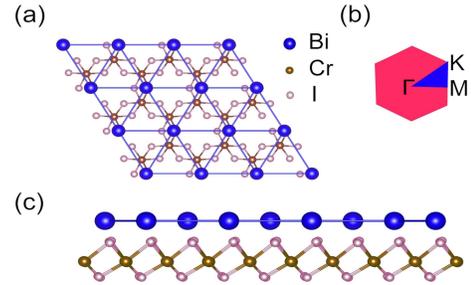}
  \caption{(a) Crystal structure of Bi monolayer on monolayer CrI$_3$ (labeled as CrI$_{3}$/Bi), and Bi atoms (blue spheres) form a triangular lattice over the substrate; (b) First Brillouin zone of CrI$_{3}$/Bi along high symmetry lines; (c) Side view of CrI$_{3}$/Bi.}
  \label{fig1}
\end{figure}

In this Letter, we theoretically reported that several candidate systems CrI$_{3}$/X
(X = Bi, Sb, or As) present non-trivial magnetic topological behaviors with a sizeable bandgap, by employing the state-of-art first-principles calculation methods. Besides a robust magnetic topology with non-zero Chern numbers, notable enhancements of the ferromagnetic properties
of pristine CrI$_3$ include the local magnetic moments, Heisenberg exchange interactions, and magnetic anisotropic energy and Curie temperature after coupling with heavy-atomic layers. Furthermore, we showed that either the ferromagnetic orientation or the strain effect can be utilized to tune the topologically nontrivial band gaps.

\textit{Ferromagnetic Properties of CrI$_3$/X}---. We launched our search for stable configurations comprised of a heavy-atoms monolayer, such as Bi, Sb, and As on the single-layer ferromagnetic insulator, CrI$_3$, to realize the quantum anomalous Hall effect in these systems. Upon a careful examination, we found that deposited heavy elements on CrI$_3$ finally formed a monolayer with a triangular lattice structure, as displayed in Fig.~\ref{fig1}(a). The evaluated interlayer binding energies for CrI$_3$/Bi, CrI$_3$/Sb, and CrI$_3$/As were $-0.65$, $-0.62$, and $-0.58$ eV per unitcell, respectively. This moderate interface interaction ensures that these atomic layers do not significantly affect the atomic structure of the pristine CrI$_3$. To our surprise, we found a significant improvement in the magnetic moment of Cr in CrI$_3$/X systems, as listed in Table~\ref{table}. This is understandable, because the apparent interface interaction can change the charge distribution
in CrI$_3$ when heavy atomic layers are deposited on CrI$_3$. As a result, the magnetic moments can be modified. This is indeed the case. Table~\ref{table} displays that the charges of Cr in CrI$_3$ and CrI$_3$/Bi are 2.17 and 1.94 e, respectively, which indicate an additional 0.23
e transferring from Bi to Cr after the deposition of Bi monolayer. Same situation occurs for CrI$_3$/Sb and CrI$_3$/As systems. However, the magnitude of charge transfer decreases from Bi to As atomic layers. This is reasonable because the attraction ability of the nucleus on the valence electrons becomes weaker, and then, the probability of charge transfer is increased when we move from As to Bi in the periodic
table. Thus, the magnetic moment of Cr atoms in a CrI$_3$ single layer can be obviously enlarged by coupling with heavy atomic layers.

\renewcommand{\arraystretch}{2}
\begin{table}
  \caption{Bader charge analysis for CrI$_3$/X systems. $\triangle$Q represents the charge difference on Cr and iodine after the deposition of heavy atomic layers on CrI$_3$ monolayer. Magnetic moments of each Cr atom, Heisenberg exchange constants, and Curie temperatures are symbolized by $M$, $J$, and $T_{\rm C}$ for CrI$_3$/X systems, respectively.}
  \begin{tabular}{>{\centering}p{1.4cm}|>{\centering}p{1cm}|>{\centering}p{1cm}|>{\centering}p{1.4cm}|>{\centering}p{1.4cm}|>{\centering}p{1.4cm}}
  \hline
  \hline
  \multirow{2}{1.4cm}{$\textrm{System}$} & \multicolumn{2}{c|}{$\mathrm{\triangle Q}$} & \multicolumn{3}{c}{Magnetic Properties}\tabularnewline
  \cline{2-6} \cline{3-6} \cline{4-6} \cline{5-6} \cline{6-6}
  & $\mathrm{Cr}$  & $\mathrm{I}$  & ${M_{\rm Cr}(\mu_{B}})$  & ${J(\rm meV)}$  & ${T_{\rm C}(\rm K)}$\tabularnewline
  \hline
  $\mathrm{CrI_{3}}$  & $0$  & $0$  & $3.0$  & $2.59$  & $44$\tabularnewline
  \hline
  $\mathrm{CrI_{3}/Bi}$ & $0.23$ & $0.27$ & $3.72$ & $9.62$ & $165$\tabularnewline
  \hline
  $\mathrm{CrI_{3}/Sb}$  & $0.21$  & $0.24$  & $3.70$  & $7.10$  & $122$\tabularnewline
  \hline
  $\mathrm{CrI_{3}/As}$  & $0.11$  & $0.14$  & $3.62$  & $6.37$  & $110$\tabularnewline
  \hline
  \hline
  \end{tabular}
  \label{table}
\end{table}

By depositing heavy atomic layers, the local magnetic moments of Cr in CrI$_3$ are considerably increased. However, what about the magnetic coupling between these enhanced magnetic moments? We found that the resulting ferromagnetic coupling in CrI$_3$ can also be enhanced. From our calculations, the ferromagnetic stabilities were $\triangle E_{\rm {FM-AFM}}$=-130, -96 and -86 meV for CrI$_3$/Bi, CrI$_3$/Sb and CrI$_3$/As systems, respectively, where $\triangle E_{\rm FM-AFM}$ represents the energy difference between ferromagnetic and anti-ferromagnetic arrangements. This indicates that the ferromagnetic nature of monolayer CrI$_3$ can be further stabilized by interacting with X(=Bi,Sb,As) atomic layers. In principle, the improvement of ferromagnetic stability boosts the Heisenberg exchange constant $J$. Our estimated Heisenberg exchange constants are respectively 9.62, 7.10, and 6.37 meV for Cr$I_{3}$/Bi, CrI$_3$/Sb, and CrI$_3$/As. These are much higher than that of monolayer CrI$_3$ ($\sim2.59$ meV)~\cite{webster2018strain}. Based on the above data and the mean field theory, the estimated Curie temperatures are 165, 122, and 110 K for CrI$_3$/Bi, CrI$_3$/Sb, and CrI$_3$/As, respectively~\cite{SM}. These are greatly enhanced from $\sim44~\rm K$ of monolayer CrI$_3$. In particular, a high Curie temperature is tall-demandable for quantum anomalous Hall system to
operate at moderate temperatures in future device applications.

After determining that CrI$_3$/X systems favor the ferromagnetic ground state, we then
proceeded to investigate the magnetic anisotropy. Our results showed that $z$ axis is an easy magnetization direction for the CrI$_{3}$/Bi
system, which is approximately 5 meV (per unit-cell with two Cr atoms) lower than that of $x$ axis. This value is much higher than that
of monolayer CrI$_3$ ($\sim$ 1.5 meV). This obvious improvement in magnetic anisotropic energy arises from the strong spin-orbit coupling in CrI$_{3}$/Bi, which has been reported in various systems~\cite{ujsaghy1996spin,yi2016atomic}.

\begin{figure}
  \includegraphics[width=8cm,angle=0]{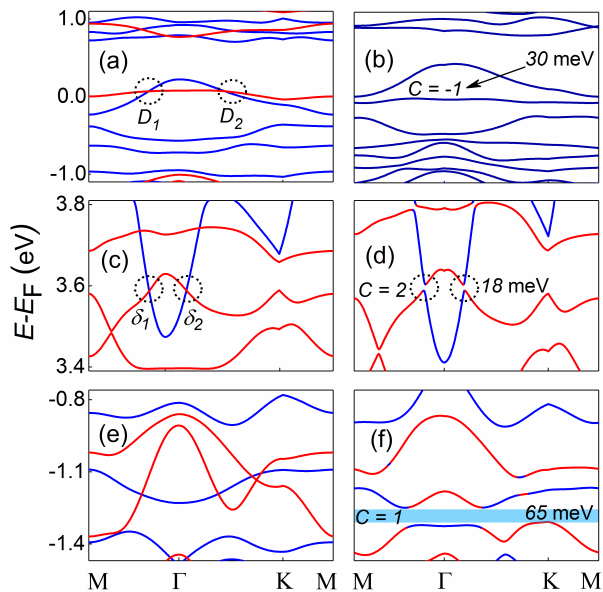}
  \caption{(a-c-e): Spin-polarized band structures of CrI$_3$/Bi (a), CrI$_3$/As (c) and CrI$_3$/Sb (e), respectively. (b-d-f): Band structures with spin-orbit coupling of CrI$_3$/Bi (b), CrI$_3$/As (d) and CrI$_3$/Sb (f), respectively. Blue and red colors denote spin-up and -down bands.}
	\label{fig2}
\end{figure}

\textit{Quantum Anomalous Hall Effect---.} The enhancement in ferromagnetic properties of pristine CrI$_3$, including the local magnetic moments, Heisenberg exchange interactions, magnetic anisotropic energy and Curie temperature, will be beneficial to realizing the quantum anomalous Hall effect in CrI$_3$ after coupling with heavy-atomic layers. Figure~\ref{fig2}(a) displays the spin-polarized band structure of CrI$_3$/Bi. The spin-up and -down bands cross around $\Gamma$ to form D$_{1/2}$ points, which are closely located near the Fermi level. Considering the presence of Bi states at the Fermi level~\cite{SM}, together with the conceptual framework of magnetic topological
insulators, one may expect the formation of topological non-trivial insulator around these avoided band-crossing points. When spin-orbit coupling is further considered in CrI$_{3}$/Bi, the degenerate states at the avoided band-crossing points repel each other to open up an energy gap (see Fig.~\ref{fig2}(b)). Our achieved direct and indirect (global) bandgaps in the CrI$_3$/Bi are respectively 30 and 75 meV. We also confirmed the non-trivial behavior inside the band gap by calculating the Chern number using the approach implemented in Wannier90 and WannierTools packages~\cite{mostofi2008wannier90,wu2018wanniertools}. Our calculations demonstrate that CrI$_{3}$/Bi indeed harbours a quantum anomalous Hall effect associated with a nonzero Chern number of $\mathcal{C}=-1$~\cite{thouless1982quantized}.

Figure~\ref{fig2}(c) displays the spin-polarized band structure for CrI$_3$/As. We found that the spin-up band crosses the spin-down bands twice within the energy window from 3.4 to 3.7 eV around $\Gamma$ point. These band-crossing points are labeled as $\delta_{1/2}$. Interestingly, the band structures around these crossing points resemble the ideal cone (Dirac point) in graphene. Due to the relatively strong spin-orbit coupling of As, one may expect a non-trivial bandgap to open around these Dirac points when spin-orbit coupling is invoked. Figure~\ref{fig2}(d) presents the band structure in the presence of spin-orbit coupling, where bandgaps open up around the crossing points of $\delta_{1/2}$, which clearly
indicate the existence of non-trivial band topologies in the conduction bands of CrI$_{3}$/As. By using the above mentioned approach, we obtained the Chern number of $\mathcal{C}=2$ for the corresponding opened band gaps. Additionally, the two valence
bands touch each other around K point~\cite{SM}. When the spin-orbit coupling is switched on, bandgaps are opened at these touching points, which signifies the topological nature of the CrI$_3$/As in valence bands. In principle, the non-trivial local bandgap in the conduction bands of CrI$_3$/As can be converted into an insulator by charge doping technique, which is equivalent to the electrostatic gating in experiment. To tune the Fermi level into the conduction band, we applied a homogenous background charge to electrostatically control the electron doping using the self-consistent simulation approach. We noticed that the Fermi level was correctly located into the spin-orbit coupling induced bandgap without affecting the band structure pattern when the homogeneous charge of $\triangle Q$=1e was functionalized~\cite{SM}. Notably, the non-trivial bandgap is also widened from 18 to 37 meV upon the functionalization of electrostatic doping~\cite{SM}. Thus, CrI$_{3}$/As can also become a quantum anomalous Hall insulator upon moderate electrostatic gating.

Figure~\ref{fig2}(e) displays the spin-polarized band structure of CrI$_{3}$/Sb. The spin-up and -down bands cross each other multiple times in the energy range between -1.30 and -1.15 eV. A non-trivial bandgap opens at these crossing points upon the application of spin-orbit coupling [see Fig.~\ref{fig2}(f)]. The global bandgap is approximately 65 meV, and exhibits a non-trivial topological nature with a Chern number of $\mathcal{C}=1$. Similar electrostatic gating approach can be used to tune the Fermi level into the bandgap.

\begin{figure*}
  \includegraphics[width=16cm,angle=0]{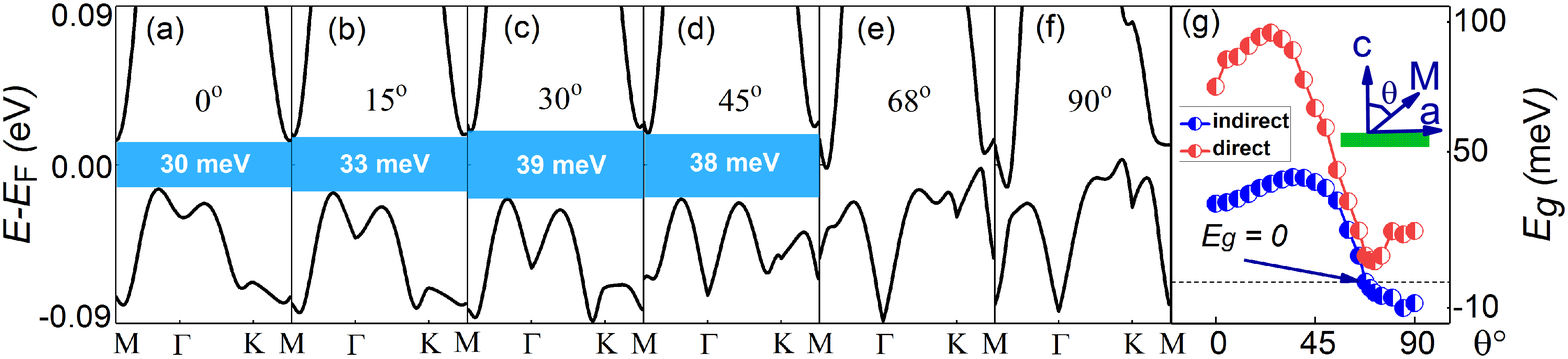}
  \caption{Effect of spin orientation on the band structure of the $\mathrm{CrI_{3}/Bi}$	system with SOC effects included. (a)-(f) present the SOC-included	band structure when all spins are aligned at angles of $0^\circ$,	$15^\circ$, $30^\circ$, $45^\circ$, $68^\circ$,	and $90^\circ$, respectively. (g) The variation in the direct and indirect bandgaps in $\mathrm{CrI_{3}/Bi}$ as a function of $\theta$, where angle $\theta$ is measured in degrees from the vertical axis. The blue and red curves correspond to the indirect and direct bandgaps, respectively.}
  \label{figure3}
\end{figure*}

\textit{Spin Orientation Effect---.} The orientation of spin moments along different crystallographic directions provides one way to control the band structure and its associated topologically non-trivial properties. In a broader context, a strong spin-orbit coupling, together with a less symmetric crystallographic structure and a suitable magnetic behavior, is the main parameter to explore the spin orientation effect. Fortunately, 2D CrI$_{3}$/X systems fulfill all these criteria, and hence, are promising candidate platforms for demonstrating the spin orientation effects. In Fig.~\ref{figure3}, our calculations show that tuning the spin moment from $c$ to $a$ axis in CrI$_{3}$/Bi system remarkably modifies the band structures, including substantially changing the Fermi surface and bandgaps. In particular, one can observe that when all spin moments were adjusted in a ferromagnetic pattern from 0$^\circ$ to $^68\circ$, the bandgap first gradually increases and then quickly decreases to be closed. Beyond $\theta=68\circ$, the CrI$_3$/Bi system undergos a topological phase transition from quantum anomalous Hall effect to a magnetic metallic state.

\begin{figure}
  \includegraphics[width=8.6cm,angle=0]{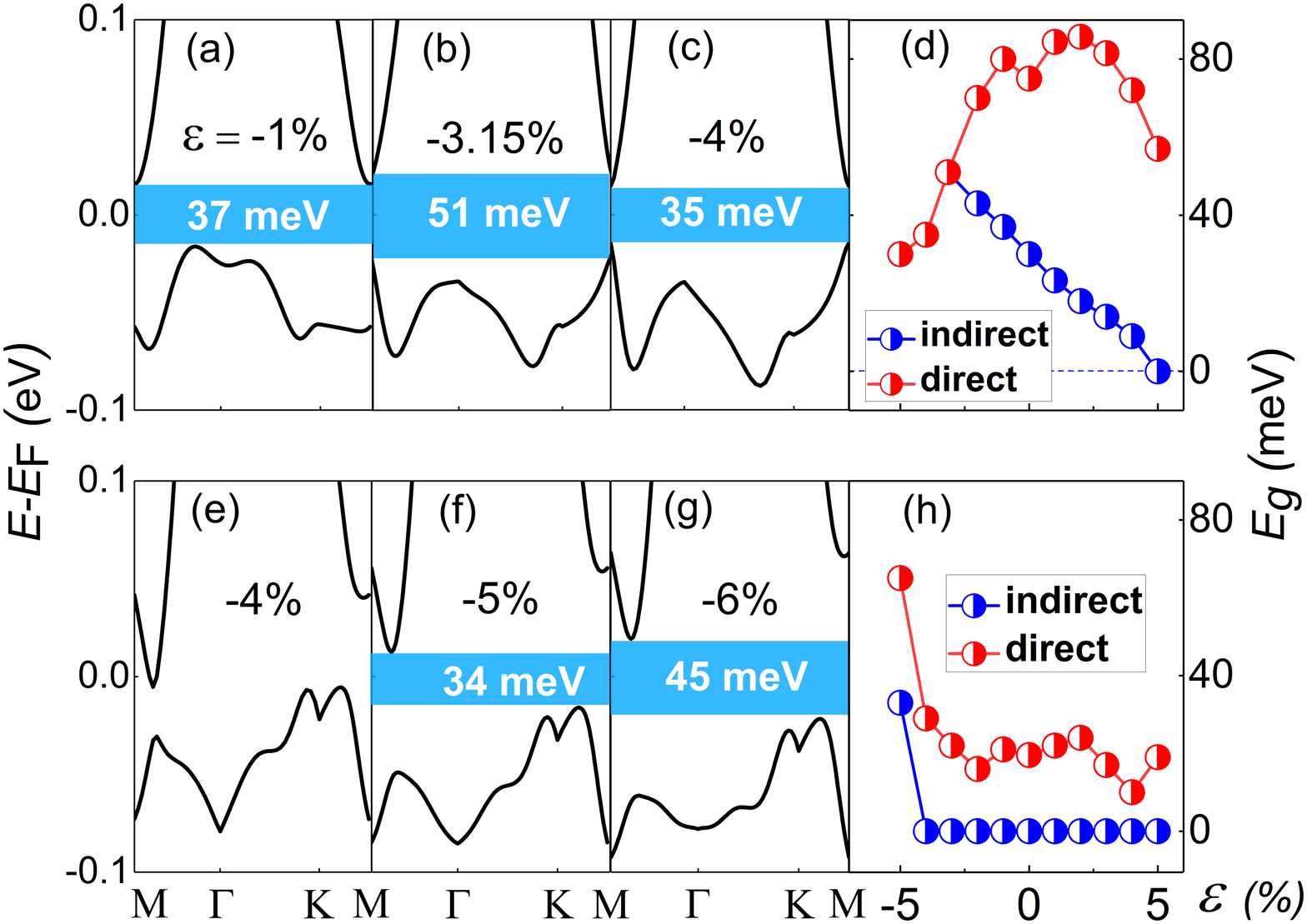}
  \caption{(a)-(d) Band structures with out-of-plane magnetization in the presence of compressive strains, i.e., -1\% (a), -3.15\% (b), and -4\% (c); (d) Bandgaps (direct and indirect) as function of compressive strain. (e)-(h) Band structures with in-plane magnetization in the presence of compressive strains, i.e., -4\% (e), -5\% (f), and -6\% (g); (h) Bandgaps (direct and indirect) as function of compressive strain. $\varepsilon$ symbolizes the strain in $\%.$}
  \label{fig4}
\end{figure}

\textit{Strain Effect---.} Strain is one of the significant approaches for engineering electronic and topological properties. Fortunately, the softness level of monolayer CrI$_3$ is higher than that of graphene and MoS$_2$, indicating that monolayer CrI$_3$ can be easily stretched or compressed experimentally~\cite{zheng2018tunable}. From this perspective, we examined the CrI$_{3}$/Bi system under the application of several biaxial strains ranging from -5 to 5\% (see Fig.~\ref{fig4}). We observed that both direct and indirect bandgaps are much sensitive to the strain. In the upper panels, we considered the out-of-plane magnetization along the vertical direction with $\theta=0^\circ$. It is observed that the global bandgap considerably increases when the lattice is allowed to compress within the horizontal plane. Interestingly, the nature of the global bandgap changes from indirect to direct at an approximately 3.15\% compressive strain (see Fig.~\ref{fig4}). While for the in-plane magnetization with $\theta=90^\circ$ [see lower panels], one can find the the global bandgap remains closed in the typical range of tensile strains. Similarly, the global bandgap remains zero against the compressive strain. However, surprisingly, the zero nature of global bandgap converts into a non-zero one when the compressive strain is increased beyond 4\%, as shown in Fig.~\ref{fig4}(h). This bandgap progressively enlarges upon a further increase in compressive strain. Therefore, in CrI$_{3}$/Bi, the quantum anomalous Hall effect with a Chern number of $\mathcal{C}=1$ can also be realized even in the presence of in-plane magnetization at a compressive strain slightly larger than 4\%.

\textit{Substrate Prototype---.} Finally, we considered CrI$_{3}$/Bi as a representative example to search for an experimentally suitable non-magnetic substrate that guarantees the survival of Dirac points around the Fermi level. We found that MoTe$_2$~(001) can serve as a good non-magnetic substrate~\cite{SM}. Our projected principles for discovering quantum anomalous Hall materials can be straightforwardly extended to other vdW 2D ferromagnetic systems. Because of the epitaxial techniques, it becomes much more convenient to artificially enlarge the family of 2D vdW ferromagnetic materials. Compared to traditional spintronic devices that are comprised of magnetic materials with some finite thickness, vdW ferromagnetic materials can simplify the device structure and shrink the device dimensions.

\textit{Summary---.} Based on density functional theory, we confirmed that CrI$_3$/Bi system is a quantum anomalous Hall insulator with a direct bandgap of approximately 30 meV at the Fermi level. We observed that CrI$_{3}$/As and CrI$_{3}$/Sb can open up topologically nontrivial band gaps in their conduction and valence bands, which can be properly tuned close the Fermi level via moderate electrostatic gating. Apart from the stable topology, we also found a major development in the ferromagnetic properties of CrI$_{3}$, including local magnetic moments, Heisenberg exchange interactions, magnetic anisotropic energy, and Curie temperatures by coupling with atomic layers of Bi/As/Sb. Furthermore, we showed that the band gaps of our proposed systems can be efficiently manipulated via tuning spin orientation or applying strain. Our study of CrI$_{3}$/X system can open new routes to explore quantum anomalous Hall effect based on 2D ferromagnetic insulators together with the potential application of CrI$_3$-based materials in spintronics.

\textit{Acknowledgements---.} This work was financially supported by the National Key R \& D Program (2017YFB0405703), the NNSFC (11474265), and Anhui Initiative in Quantum Information Technologies. We also thank the supercomputing service of AMHPC and the Supercomputing Center of USTC for providing the high-performance computing resources.


\begin{thebibliography}{99}
\bibitem{weng2015quantum}
 H. Weng, R. Yu, X. Hu, X. Dai, and Z. Fang, Adv. Phys. \textbf{64}, 227 (2015)	
\bibitem{ren2016topological} 	
 Y. Ren, Z. Qiao, and Q. Niu, Rep. Prog. Phys. \textbf{79}, 066501 (2016).
 \bibitem{he2018topological}
 K. He, Y. Wang, and Q.-K. Xue, Annu. Rev. Condens. Matter Phys. \textbf{9}, 329 (2018).
 \bibitem{liu2018intrinsic}
 Z. Liu, G. Zhao, B. Liu, Z. Wang, J. Yang, and F. Liu, Phys. Rev. Lett. \textbf{121}, 246401 (2018).
 \bibitem{ni2018intrinsic}
 X. Ni, W. Jiang, H. Huang, K.-H. Jin, and F. Liu, Nanoscale \textbf{10}, 11901 (2018).
 \bibitem{zanolli2018hybrid}
 Z. Zanolli, C. Niu, G. Bihlmayer, Y. Mokrousov, P. Mavropoulos, M. Verstraete, and S. Blugel, Phys. Rev. B \textbf{98}, 155404 (2018).
 \bibitem{ren2018quantum}
 Y. Ren, T.-S. Zeng, W. Zhu, and D. Sheng, Phys. Rev. B \textbf{98}, 205146 (2018).
 \bibitem{deng2018quantum}
 X. Deng, H. Yang, S. Qi, X. Xu, and Z. Qiao, Front. Phys. \textbf{13}, 137308 (2018).
 \bibitem{tse2011quantum}
 W.-K. Tse, Z. Qiao, Y. Yao, A. MacDonald, and Q. Niu, Phys. Rev. B \textbf{83}, 155447 (2011).
 \bibitem{qiao2010quantum}
 Z. Qiao, S. A. Yang, W. Feng, W.-K. Tse, J. Ding, Y. Yao, J. Wang, and Q. Niu, Phys. Rev. B \textbf{82}, 161414 (2010).
 \bibitem{ding2011engineering}
 J. Ding, Z. Qiao, W. Feng, Y. Yao, and Q. Niu, Phys. Rev. B \textbf{84}, 195444 (2011).
 \bibitem{zhang2012electrically}
H. Zhang, C. Lazo, S. Blugel, S. Heinze, and Y. Mokrousov, Phys. Rev. Lett. \textbf{108}, 056802 (2012).
\bibitem{qiao2012microscopic}
 Z. Qiao, H. Jiang, X. Li, Y. Yao, and Q. Niu, Phys. Rev. B \textbf{85}, 115439 (2012).
 \bibitem{jiang2012quantum}
H. Jiang, Z. Qiao, H. Liu, and Q. Niu, Phys. Rev. B \textbf{85}, 045445 (2012).
 \bibitem{kou2013interplay}
X. Kou, M. Lang, Y. Fan, Y. Jiang, T. Nie, J. Zhang, W. Jiang, Y. Wang, Y. Yao, L. He, et al., Acs Nano \textbf{7}, 9205 (2013).
 \bibitem{pan2014valley}
H. Pan, Z. Li, C.-C. Liu, G. Zhu, Z. Qiao, and Y. Yao, Phys. Rev. Lett. \textbf{112}, 106802 (2014).
 \bibitem{qi2016high}
S. Qi, Z. Qiao, X. Deng, E. D. Cubuk, H. Chen, W. Zhu, E. Kaxiras, S. Zhang, X. Xu, and Z. Zhang, Phys. Rev. Lett. \textbf{117}, 056804 (2016).
\bibitem{deng2017realization}
X. Deng, S. Qi, Y. Han, K. Zhang, X. Xu, and Z. Qiao, Phys. Rev. B \textbf{95}, 121410 (2017).
\bibitem{jin2015quantum}
 K.-H. Jin and S.-H. Jhi, Sci. Rep. \textbf{5}, 8426 (2015).
 \bibitem{hsu2017quantum}
 C.-H. Hsu, Y. Fang, S. Wu, Z.-Q. Huang, C. P. Crisostomo, Y.-M. Gu, Z.-Z. Zhu, H. Lin, A. Bansil, F.-C. Chuang, et al., Phys. Rev. B \textbf{96}, 165426 (2017).
 \bibitem{luo2013massive}
 W. Luo and X.-L. Qi, Phys. Rev. B \textbf{87}, 085431 (2013).
 \bibitem{yang2013proximity}
 H.-X. Yang, A. Hallal, D. Terrade, X. Waintal, S. Roche, and M. Chshiev, Phys. Rev. Lett. \textbf{110}, 046603 (2013).
\bibitem{qiao2014quantum}
Z. Qiao, W. Ren, H. Chen, L. Bellaiche, Z. Zhang, A. MacDonald, and Q. Niu, Phys. Rev. Lett. \textbf{112}, 116404 (2014).
\bibitem{xu2015quantum}
G. Xu, J. Wang, C. Felser, X.-L. Qi, and S.-C. Zhang, Nano Lett. \textbf{15}, 2019 (2015).
\bibitem{zhang2018strong}
J. Zhang, B. Zhao, T. Zhou, Y. Xue, C. Ma, and Z. Yang, Phys. Rev. B \textbf{97}, 085401 (2018).
\bibitem{chang2013experimental}
C.-Z. Chang, J. Zhang, X. Feng, J. Shen, Z. Zhang, M. Guo, K. Li, Y. Ou, P. Wei, L.-L. Wang, et al., Science \textbf{340}, 167 (2013).
\bibitem{liu2016quantum}
C.-X. Liu, S.-C. Zhang, and X.-L. Qi, Annu Rev. Condens. Matter Phys. \textbf{7}, 301 (2016).
\bibitem{bestwick2015precise}
A. Bestwick, E. Fox, X. Kou, L. Pan, K. L. Wang, and D. Goldhaber-Gordon, Phys. Rev. Lett. \textbf{114}, 187201 (2015).

\bibitem{MnBiTe}
Y. Deng, Y. Yu, M. Z. Shi, J. Wang, X. H. Chen, and Y. Zhang, arXiv:1904.11468 (2019).

\bibitem{garrity2013chern}
K. F. Garrity and D. Vanderbilt, Phys. Rev. Lett. \textbf{110}, 116802 (2013).


\bibitem{webster2018strain}
L. Webster and J.-A. Yan, Phys. Rev. B \textbf{98}, 144411 (2018).

\bibitem{SM}
See details in Supplemental Materials.


\bibitem{ujsaghy1996spin}
O. \'Ujs\'aghy, A. Zawadowski, and B. L. Gyorffy, Phys. Rev. Lett. \textbf{76}, 2378 (1996).
\bibitem{yi2016atomic}
D. Yi, J. Liu, S.-L. Hsu, L. Zhang, Y. Choi, J.-W. Kim, Z. Chen, J. D. Clarkson, C. R. Serrao, E. Arenholz, et al., PNAS \textbf{113}, 6397 (2016).
\bibitem{mostofi2008wannier90}
A. A. Mostofi, J. R. Yates, Y.-S. Lee, I. Souza, D. Vanderbilt, and N. Marzari, Comput. Phys. Commun. \textbf{178}, 685 (2008).
\bibitem{wu2018wanniertools}
Q. Wu, S. Zhang, H.-F. Song, M. Troyer, and A. A. Soluyanov, Comput. Phys. Commun. \textbf{224}, 405 (2018).
\bibitem{thouless1982quantized}
D. J. Thouless, M. Kohmoto, M. P. Nightingale, and M. den Nijs, Phys. Rev. Lett. \textbf{49}, 405 (1982).
\bibitem{zheng2018tunable}
F. Zheng, J. Zhao, Z. Liu, M. Li, M. Zhou, S. Zhang, and P. Zhang, Nanoscale \textbf{10}, 14298 (2018).
\end{thebibliography}
\end{document}